\documentclass{aa}

\usepackage[utf8]{inputenc}
\usepackage{newunicodechar}
\newunicodechar{∼}{\textasciitilde{}}
\usepackage{graphicx}
\usepackage{txfonts}
\usepackage{natbib}
\usepackage{mathrsfs}
\usepackage{hyperref} 
\usepackage{dsfont}
\hypersetup{colorlinks=true,linkcolor=blue,citecolor=blue,urlcolor=blue}
\usepackage{orcidlink}
\usepackage{lscape}

\begin{document}







\title{The LEGARE Project. I.}
\subtitle{Chemical evolution model of the  Nuclear Stellar Disc in a Bayesian framework}

\author { E. Spitoni \orcidlink{0000-0001-9715-5727}\inst{1,2}   \thanks {email to: emanuele.spitoni@inaf.it},  M. Schultheis \orcidlink{0000-0002-6590-1657}  \inst{3}, F. Matteucci \orcidlink{0000-0001-7067-2302}  \inst{1,4,5}, N. Ryde  \orcidlink{0000-0001-6294-3790} \inst{6}, G. Cescutti \orcidlink{0000-0002-3184-9918} \inst{4,5,1}, A. Saro  \orcidlink{0000-0002-9288-862X} \inst{4,2,1,5,7}, \\
M.C. Sormani \orcidlink{0000-0001-6113-6241} \inst{8} \and B. Thorsbro \orcidlink{0000-0002-5633-4400}\inst{3,6}}
\institute{I.N.A.F. Osservatorio Astronomico di Trieste, via G.B. Tiepolo
 11, 34143, Trieste, Italy  
  \and IFPU, Institute for Fundamental Physics of the Universe, Via Beirut 2, I-34151 Trieste, Italy  
  \and Université Côte d’Azur, Observatoire de la Côte d’Azur, Laboratoire Lagrange, CNRS, Blvd de l’Observatoire, 06304 Nice,
France
  \and Dipartimento di Fisica, Sezione di Astronomia, Università di Trieste, Via G. B. Tiepolo 11, 34143 Trieste, Italy
 \and  INFN Sezione di Trieste, via Valerio 2, 34134 Trieste, Italy
 \and Division of Astrophysics, Department of Physics, Lund University, Box 118, 221 00 Lund, Sweden
 \and ICSC – Italian Research Center on High Performance Computing, Big Data and Quantum Computing, Via Magnanelli 2,
40033 Casalecchio di Reno, Italy
 \and Como Lake centre for AstroPhysics (CLAP), DiSAT, Università
dell’Insubria, via Valleggio 11, Como, 22100, Italy
  }

 \date{Received xxxx / Accepted xxxx}

\abstract {The Nuclear Stellar Disc (NSD) of the Milky Way is a dense, rotating stellar system in the central ∼200 pc. The NSD is thought to be primarily fuelled by bar-driven gas inflows from the inner Galactic disc. 
}
{As part of  the LEGARE project, we aim to construct the first chemical evolution models for the NSD using a Bayesian approach tailored to reproduce the observed metallicity distribution functions  and  compared with the available abundance  ratios  for Mg, Si, Ca relative to Fe. In particular, we test whether the flowing gas  from the inner Galactic disc, which feeds the  NSD, can reproduce the observed abundances.}
{We adopt a state-of-the-art chemical evolution model in which the gas responsible for the formation of the NSD is assumed to be driven by  the Galactic bar-induced inflows.
The chemical composition of the accreted material is assumed to reflect that of the Galactic disc at a radius of $\sim$4 kpc.
A Bayesian framework based on MCMC techniques is then employed to fit  the metallicity distribution functions  of different  samples of    NSD stars.} 
{  If we take the NSD data at face value, without considering a possible contamination from bulge stars, we find that a formation scenario based  on the inner disc flowing gas is inconsistent with the low metallicity tail of the observed metallicity distribution function. This is because the inner disc metallicity, at the epoch of bar formation, was already near solar. On the other hand, models invoking dilution from additional metal-poor inflows successfully reproduce the observations.
 Models with different  levels of gas dilution, share similar gas infall timescales (ranging from 3.7 to 5.2 Gyr) and negligible galactic winds (mass-loading factors $\omega$ between 0.001 and 0.030). The best-fit model corresponds to an inflow with a metallicity 5 times lower than that of the inner disc, and a moderate star-formation efficiency. The same model successfully reproduces the observed [$\alpha$/Fe]–[Fe/H] abundance trends and predicts a star formation history consistent with the most recent estimates.  However, if we assume that the metallicity distribution function is contaminated by metal poor bulge stars and is restricted to stars with [Fe/H] > –0.3 dex, there is no more need for gas dilution. In this case, the best-fit model is characterised by a very low star formation efficiency, coupled with a mild galactic wind.} 
 { Our analysis indicates that dilution of the inflowing gas forming the NSD, is necessary to reproduce its observed chemical properties, if bulge contamination in the data is not considered. This implies that, in addition to bar-driven inflows from the inner thin disc, lower metallicity gas - possibly originating from the thick disc or from more recent accretion events - contributed to the formation of the NSD. On the other hand, when contamination by bulge stars is assumed, dilution is no more required.}

\keywords{Galaxy: disk -- Galaxy: abundances -- Galaxy: formation -- Galaxy: evolution  -- ISM: abundances}

\titlerunning{CEMs of the NSD}

\authorrunning{Spitoni et al.}

\maketitle

\section{Introduction}

The central regions of the Milky Way provide a unique laboratory to study the interplay between gas dynamics, star formation, and chemical enrichment. The NSD is a significant component within the central region of
our Milky Way, alongside the nuclear star cluster (NSC) and the central massive black hole (see \citealt{schultheis2025} for a review and references therein).
Among these structures, the NSD stands out as a massive, rotating component confined within the central few hundred parsecs, with a scale radius of $\sim 100$ pc and a scale height of around 50 pc characterised by a stellar mass of the order of $10^9$ M$_\odot$ \citep{Launhardt2002, NoguerasLara2020}. Its high stellar density, rapid rotation, and distinct geometry relative to the surrounding bulge make the NSD a key tracer of the secular evolution of the Galaxy.

The stellar
composition of the NSD has been relatively unexplored until recent times, primarily due to challenges such
as extreme extinction and stellar crowding. NSDs are also seen in extragalactic systems and show an inside-
out formation \citep[e.g.,][]{bittner2020}. The Galactic bar is a structure that cannot be ignored if we want to
describe properly the evolution of the NSD. Simulations suggest that most of the NSD mass forms within $\sim$1
Gyr of the bar formation \citep{baba2020,cole2014}. Hence, the NSD star formation history
can be used to estimate the age of the Galactic bar. Similar to most disc galaxies, the Milky Way features a
central Galactic bar that dominates approximately the inner 4 kpc. As the bar sweeps through the gas in the
Galactic disc, it creates long shock lanes along its leading edges and tips \citep{athanassoula1992}. This gas falls
on the leading side of the bar into the central region and accumulates on the dense nuclear disc \citep{contopoulos1989}. In the recent work of  \citet{nieuwmunster2024}, they presented a detailed orbital
analysis of stars located in the NSD of the Milky Way, investigating also the dynamical history of this
structure. Based on a detailed orbital analysis, they were able to classify orbits into various families, most of
which are characterised by x2-type orbits, which are dominant in the inner part of the bar. This is another
confirmation that the NSD evolution is strictly linked with the destiny of the Galactic bar.

Recent observational progress has brought the NSD into sharper focus. Using VLT/KMOS, \citet{Schultheis2021} demonstrated that the NSD is chemically and kinematically distinct from both the Galactic bulge and the NSC. Metal-rich stars display a dynamically cool component with lower velocity dispersion, likely formed from gas in the Central Molecular Zone (CMZ), while the origin of the more metal-poor population remains debated. The GALACTICNUCLEUS survey has also provided exquisite photometric constraints, revealing a complex star formation history with evidence for both old and intermediate-age populations \citep{NoguerasLara2021}. 
At the chemical level, recent high-resolution abundance studies have delivered unprecedented detail. \citet{Ryde2025} used IGRINS/Gemini South to measure abundances of 18 elements in nine NSD M giants, covering $-1.0 \lesssim {\rm [Fe/H]} \lesssim +0.5$. Their results show strong similarities with inner-disc and inner-bulge populations at sub-solar metallicity, while at super-solar metallicity the NSD follows the trends of the NSC.  These results demonstrate that high-quality chemical abundances can be obtained even in the heavily extincted Galactic centre, opening the way for   large-scale spectroscopic programs such as  the Multi-Object Optical and Near-infrared Spectrograph   \citep[MOONS,][]{cirasuolo2020}.

Despite this progress, dedicated chemical evolution models of the NSD remain scarce. While many studies have successfully modeled the bulge in different regions  \citep[e.g.][]{matteucci2019, nieuw2023}, Galactic center \citep{grieco2015,thorsbro2020} the NSD’s unique environment—shaped by bar-driven gas inflows, CMZ dynamics, and intense star formation—requires a tailored approach. In particular, \citet{friske2025} have recently attempted to model  the chemo-dynamical evolution of the NSD.  While their models provide valuable insights into the role of gas inflows and dynamical processes, their predicted abundance-ratio in the gas phase  trends display significantly more irregular behaviour than what the \citet{Ryde2025}  which show smooth [$\alpha$/Fe] versus [Fe/H] sequences at more extended metallicities. Moreover, no comparison has been made with the available metallicity distributions of NSD stars.

As part of the LEGARE (Linking the chemical Evolution of Galactic discs AcRoss diversE scales) project, we present a new chemical evolution model for the NSD, developed within a Bayesian framework. Our approach aims to directly reproduce the metallicity distribution function (MDF) of a subsample from the \citet{Schultheis2021}  analysis. This allows us to quantitatively test different scenarios for the chemical composition of the gas flows responsible for the NSD's formation and to derive robust posterior distributions for the model's key parameters. We assume the inflowing gas's composition is driven by bar-induced inflows from the inner galactic disc, consistent with the formation of the Galactic bar approximately 8 Gyr ago, as described by \citep{sanders2024}.
This approach enables us to robustly quantify parameter degeneracies and uncertainties within our NSD model, a crucial step for building a reliable and data-driven analysis. The adopted methodology closely follows the framework introduced by \citet{Spitoni2020,spitoni2021}, who applied MCMC techniques to explore the parameter space of Galactic disc chemical evolution models.

Our paper is organised as follows: in Section~\ref{models_sec}, we present the chemical evolution models for the inner Galactic disc and the NSD. In Section \ref{sec_data_MDF}, the data sample used the MCMC analysis is described. The fitting procedure is outlined in Section \ref{fitting}. Finally, our model results and conclusions are presented in Sections  \ref{sec_results} and \ref{conclu_sec}, respectively.


\begin{figure*}
 \centering
  \includegraphics[scale=0.54]{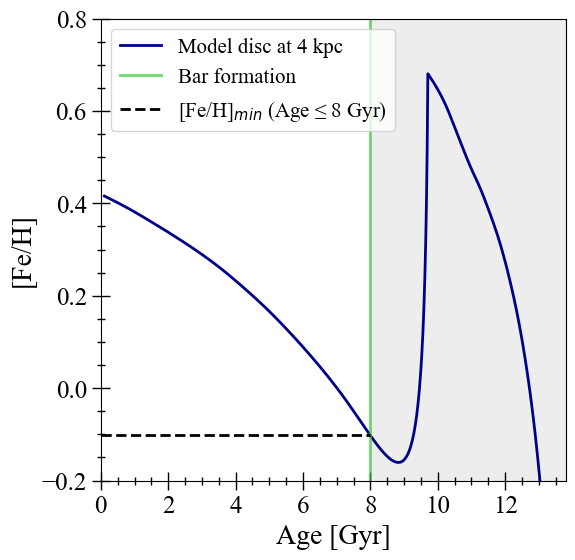}
  \includegraphics[scale=0.38]{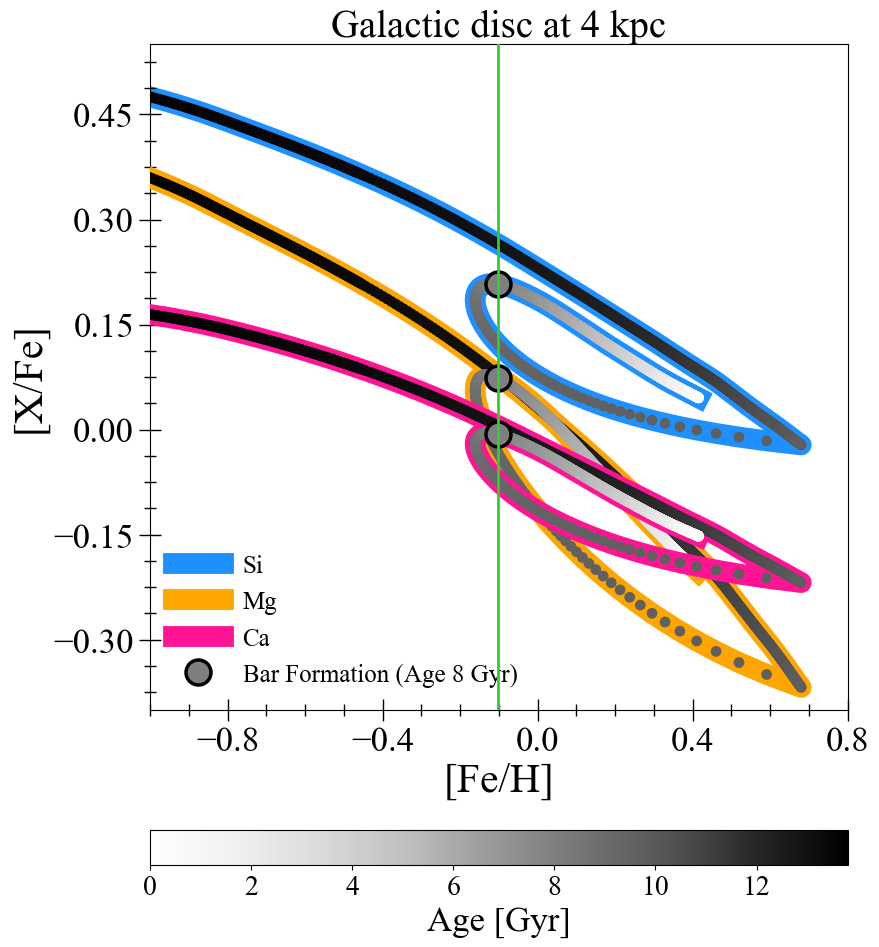} 
  \caption{Chemical evolution of the inner Galactic disc at 4 kpc, computed with the same model parameters adopted in \citet{spitoni2021} but using the stellar yields of \citet{romano2010}. {\it Left panel}: Predicted age-metallicity relation highlighting the minimum [Fe/H] value after the bar formation (Age $\leq$ 8 Gyr).
{\it Right panel}: Evolution of  the abundance ratios  [Fe/H] versus [X/Fe]  (with X = Mg, Si, Ca).
  Large black-edged circles connected by the vertical green line indicate the predicted chemical composition of the ISM at the epoch of bar formation.
  }
\label{fig_disc}
\end{figure*}

\begin{figure}
 \centering
       \includegraphics[scale=0.38]{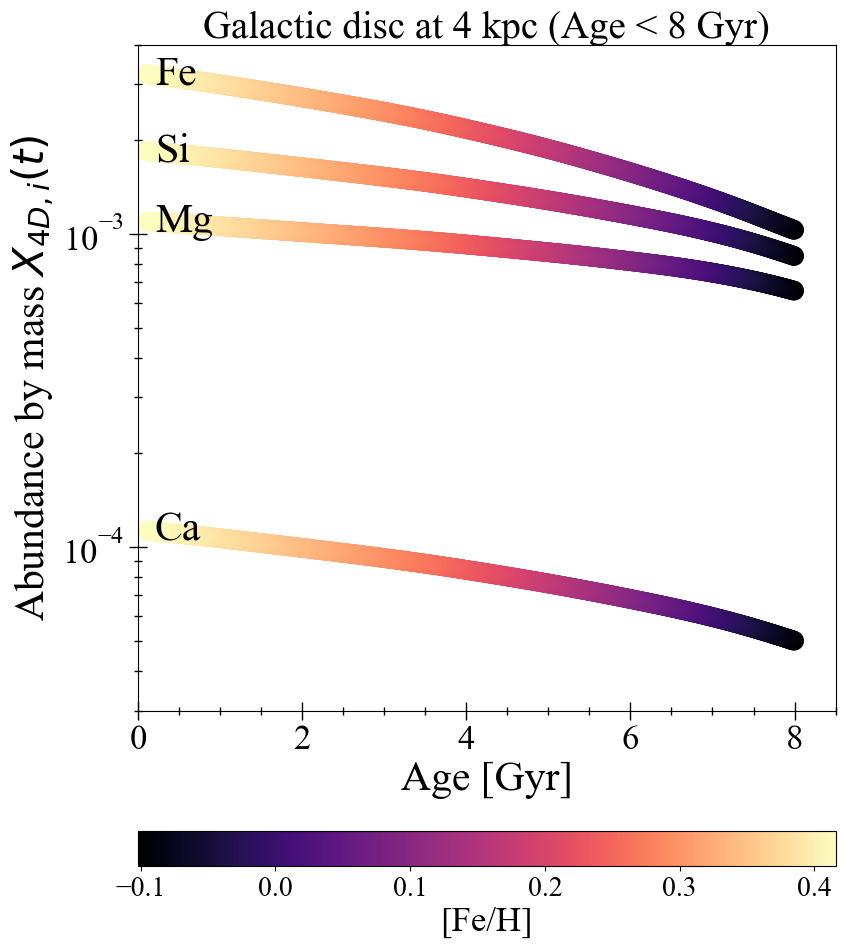}
 \caption{Evolution of the abundance by mass $X_{4D,i}(t)$ of the  elements considered in this study ($i$=Mg, Si, Ca and Fe)  as predicted by the model of the inner Galactic disc at 4 kpc  after the formation of the Galactic bar (Age< 8 Gyr). }
\label{fractionX}
\end{figure}

\section{Chemical evolution models}
\label{models_sec}

In this Section, we present the chemical evolution models adopted in this study. 
Section \ref{model_4kpc} describes the chemical evolution of the inner Galactic disc at a radius of 4 kpc, following the framework of \citet{spitoni2021}, 
while Section \ref{NSD_sec} introduces the details of the new model developed for the NSD.

\subsection{Chemical evolution model for the inner Galactic disc}
\label{model_4kpc}
In \citet{spitoni2021}, the authors   presented a multi-zone two-infall chemical evolution model designed to reproduce  APOGEE DR16 \citep{Ahumada2019} abundance ratios at different Galactocentric distances using a Bayesian analysis based on MCMC methods. For this work we will consider the best model  for the innermost Galactic region disc computed at 4 kpc. 

To trace the evolution of the thick and thin disc components, they adopted a two-infall prescription for the gas accretion, whose functional form for the infall rate at 4 kpc is:
\begin{eqnarray}
\mathcal{I_{\it 4D, i}}(t)&=& X_{1,i} \, \Big( A \, e^{-t/ {\rm T}_{ \rm high}}  \Big)
+ X_{2,i} \Big(
 \theta(t-{\rm T}_{{\rm d}})  \, B \, e^{-(t-{\rm T}_{{\rm d}})/{\rm T}_{ \rm low}} \Big),
\label{infall}
 \end{eqnarray}
where ${\rm T}_{{\rm high}}$=0.11 Gyr, ${\rm T}_{{\rm low}}$=0.376 Gyr,     are the timescales of the two distinct gas infall episodes. 
The Heaviside step function is represented by $\theta$. 
  $X_{1,i}$ and  $X_{2,i}$ is the abundance by mass unit of the element $i$ in  the two different infalling gas.   They suggested that for the high-$\alpha$ the infall has a primordial composition whereas 
for the second gas infall
 a chemical enrichment must be obtained from the model of the high-$\alpha$ disc phase corresponding to [Fe/H]=-0.5 dex. 
 Finally, the coefficients $A$ and $B$ are associated to the surface mass densities of the disc components. The star formation rate (SFR) follows the \citet{kenni1998} law:
\begin{equation}
\psi_{4D}(t)\propto \nu_{\rm high, low} \cdot \sigma_{g}(t)^{k},
\label{k1}
\end{equation}
 where $\sigma_g$ is the gas surface
 density and $k = 1.5$ is the exponent.  
The parameter $\nu_{\rm high,low}$ denotes the star formation efficiency (SFE), which can take different values during distinct phases of Galactic evolution. In particular, \citet{spitoni2021} adopted $\nu_{\rm high}=3$ Gyr$^{-1}$ and $\nu_{\rm low}=1$ Gyr$^{-1}$. The remaining best-fit parameters obtained from the Bayesian analysis are reported in Table 2 of \citet{spitoni2021}.

Figures \ref{fig_disc} and \ref{fractionX} show  model results for the Galactic disc  at a Galactocentric distance of 4 kpc. They are obtained using the parameter set of Table 2 in \citet{spitoni2021}, the nucleosynthesis prescriptions of \citet[][see Section \ref{nucleo_sec} for details]{romano2010}, and adopting the initial stellar mass function (IMF) of \citet{kroupa1993}, assumed to be constant in both time and space for consistency with the NSD model. The adopted solar abundance values are the ones of \citet{asplund2009}.

 The left panel of Fig.~\ref{fig_disc} shows the age–metallicity relation. We  highlight that the  accretion of gas in the second infall with a chemical enrichment level significantly lower than that already achieved by the thick disc ([Fe/H] $\simeq -0.5$ dex), temporarily decreases the metallicity of the stellar populations formed immediately after the infall event 
(see discussion in \citealt{spitoni2019,spitoni2021,spitoni2022}).
We also indicated  the minimum metallicity reached after the onset of the bar, [Fe/H]$_{\mathrm{min}}=-0.10$ dex.   In the right panel of Fig. \ref{fig_disc}, we present the temporal evolution of  the abundance ratios  [Fe/H] versus [X/Fe] (with X = Mg, Si, Ca) predicted by the Galactic disc model at 4 kpc. Shortly after the onset of the second infall, a distinctive loop appears in the chemical plane, displaying a ribbon-like shape \citep{spitoni2024}. In fact, the accretion of the above-mentioned chemically poor gas  triggers a dilution phase. Once star formation resumes, CC-SNe drive a sharp increase in the [X/Fe] ratio, which is then followed by a decline and a shift toward higher metallicities, as Type Ia SNe contribute substantial amounts of Fe.

We define the abundance by mass of an element $i$ in the inner Galactic disc, computed at 4 kpc, as the quantity:
\begin{equation}
  X_{4D,i}(t)=M_{4D,i}(t)/M_{4D, \, \rm gas}(t)  
\end{equation}
Here, $M_{4D,i}(t)$ denotes the mass of element $i$ contained in the interstellar medium (ISM), while $M_{4D,\mathrm{gas}}$ represents the total ISM gas mass, both evaluated at a Galactocentric distance  of 4 kpc.
In Fig. \ref{fractionX}  we draw the  $X_{4D,i}(t)$ quantities  for Mg, Si, Ca, and Fe,  as a function of Galactic age after the formation of the bar (Age<8 Gyr), during which the NSD is expected to have formed and grown over time.

\subsection{The chemical evolution model for the NSD}
\label{NSD_sec}
As indicated in the Introduction, the NSD is primarily fueled by gas channeled inward by the Galactic bar from the inner disc regions. In this work, we refer generically to gas "flows" without distinguishing between material originating in the disc and any possible residual contribution from other Galactic components. Our focus is on the chemical enrichment  expected for the NSD, and in particular on constraining the degree of chemical pre-enrichment from the Galactic disc that is required to reproduce the available spectroscopic observations.  We emphasise that our model assumes in situ formation for all NSD stars, whereas the gas reservoir feeding their formation may be supplied by the Galactic disc or other Galactic components. Hence, the  equation of the evolution of the surface density of the chemical element $i$ is: 

\begin{equation}
\label{gas_rate}
 \dot\sigma_{NSD, \,i}= -\psi_{NSD}(t) \, {X}_{NSD,\,i}(t)+   R_{NSD, \,i}(t)+  \mathscr{F}_{NSD, \,i}(t) - W_{NSD, \,i}(t).
\end{equation} 
 The first term on the right-hand side of the equation accounts for the depletion of chemical elements from the ISM as they are locked into newly formed stars. Here,  ${X}_{NSD,i}(t)$ denotes the abundance by mass of a given element $i$, $\psi_{NSD}$ is the SFR  defined by the  \citet{kenni1998} law  of eq. (\ref{k1}) with the same exponent  $k$ as the disc model and $\nu$ the star formation efficiency. The second term,  $R_{NSD, \,i}(t)$, represents the fraction of matter that is
restored to the ISM in the form of the element $i$ originating from stellar winds and 
SN explosions.  
 The $\mathscr{F}_{NSD, \,i}$ quantity indicates  the gas flow rate   onto the NSD  for the element $i$, and it  can be written as:
\begin{equation}
  \mathscr{F}_{NSD, \,i}(t)  = {X}_{ \mathscr{F}, NSD, \, i}(t) \,
 \theta(t-{\rm T}_{{\rm bar}})   \, \mathscr{C} \, e^{-(t-{\rm T_{bar}})/{\tau}},
\end{equation} 
which enforces that the NSD can only form after the onset of the bar. The latter is assumed to have developed $\rm T_{bar}= 8$ Gyr ago, in agreement with \citet{sanders2024}. The parameter $\mathscr{C}$ is constrained by the total surface mass density of the NSD through this expression:

\begin{equation}
 \mathscr{C}=\frac{\sigma_{NSD}}
   {\displaystyle \int_{\rm T_{bar}}^{13.8 \, \rm {Gyr}}
    e^{-(t'-{\rm T_{bar}})/{\tau}} \, dt'},
\end{equation}
having imposed that the present-day total surface mass density $\sigma_{NSD}$ of the NSD is
\begin{equation}
\label{sigma}
\sigma_{NSD} = \frac{5\cdot10^9}{\pi \cdot 100^2}\sim  5\cdot10^4 \, \Big[ {\rm M}_{\odot} /{\rm pc}^{2} \Big].
\end{equation}
In eq. (\ref{sigma}), we assumed that the radius of the NSD is $R_{\rm NSD} = 100$~pc and that its present-day total mass (gas plus stars) is $5 \cdot 10^9$~M$_{\odot}$.  Finally, the quantity ${X}_{ \mathscr{F}, NSD, \, i}(t)$ represents the chemical composition of the  gas that feeds the NSD.
 In Section \ref{sec_results_cem_comp}, we will present the different scenarios for the chemical composition of the gas inflowing into the NSD explored in this study.
In addition, in eq. \ref{gas_rate},  we assume the presence of a  wind  proportional to the SFR:
\begin{equation}
    W_{NSD,i}(t)=\omega \, {X}_{NSD, \, i}(t) \, \psi_{NSD}(t),
\end{equation}
where $\omega$ is the loading factor.

\begin{figure*}
    \centering
    \includegraphics[scale=0.38]{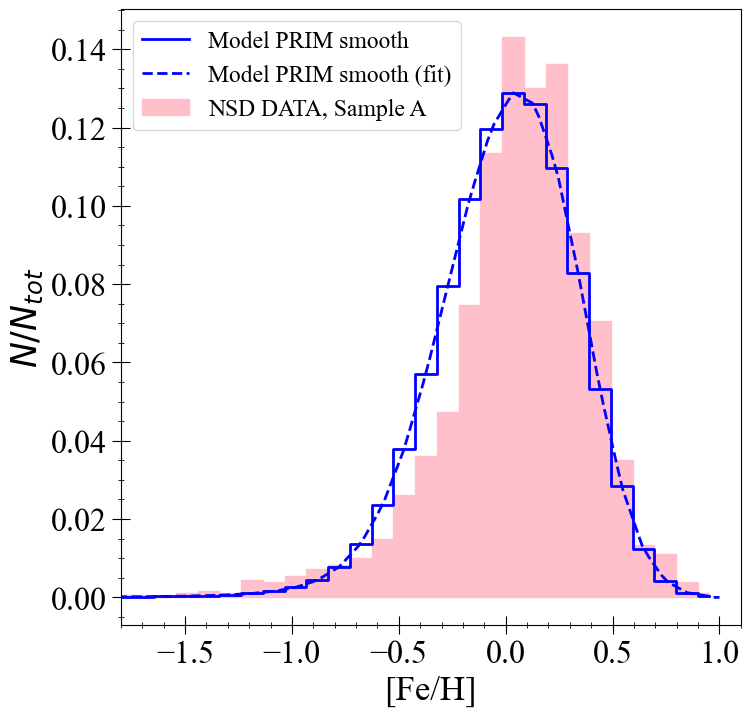}
     \includegraphics[scale=0.38]{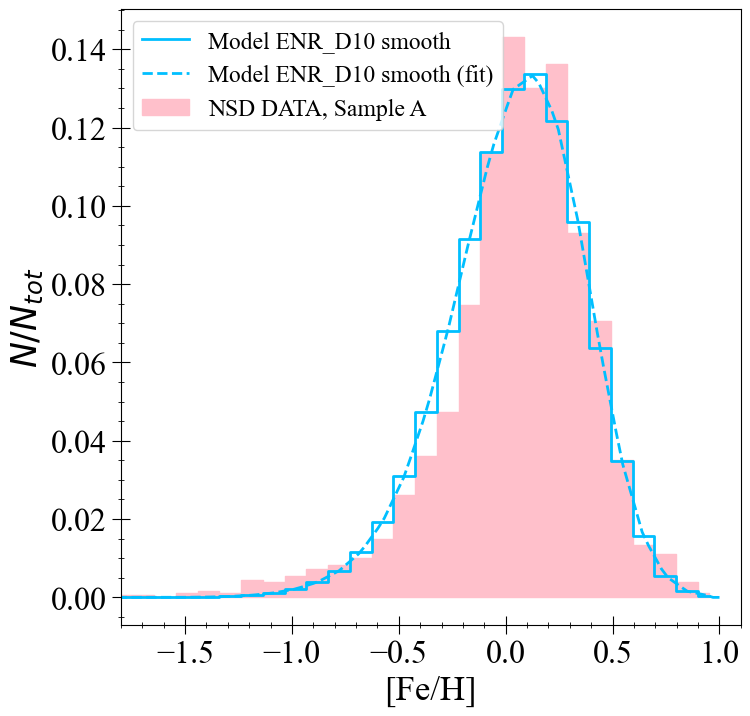}
 \includegraphics[scale=0.38]{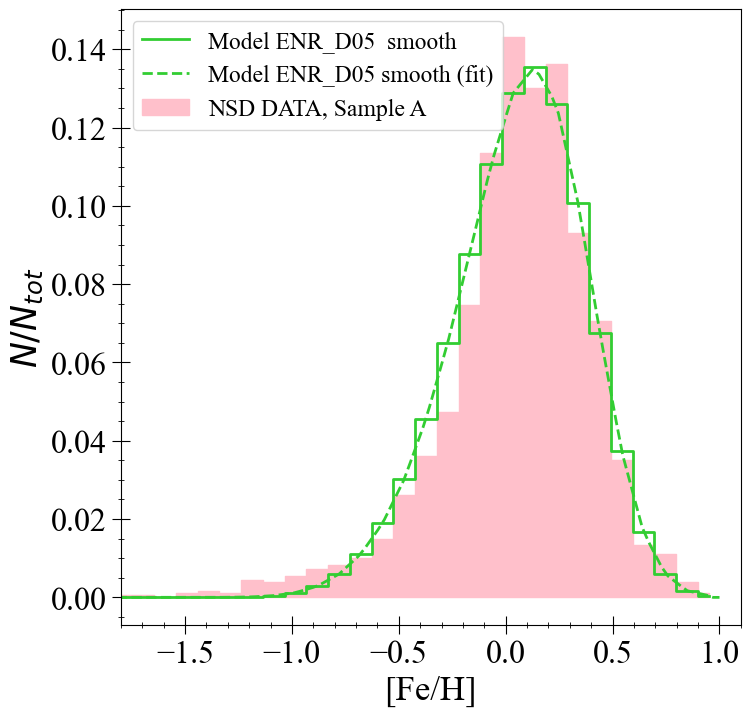}
 \includegraphics[scale=0.38]{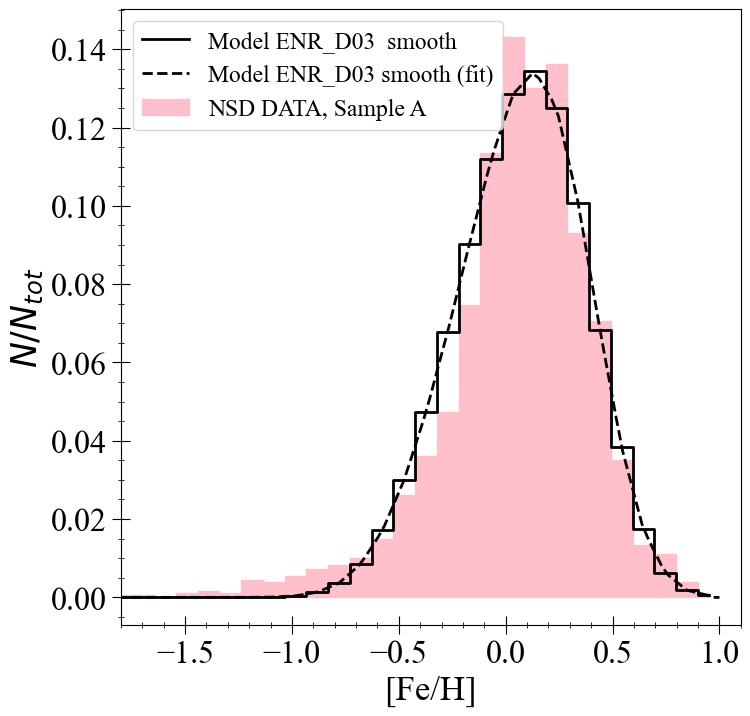}
 \caption{Comparison between the observed NSD MDF (Sample A, filled distributions, see Section \ref{sec_data_MDF} for details) and the predictions of the best-fit models (with parameters listed in Table \ref{tab_mcmc}), after applying a Gaussian convolution with $\sigma_{\rm [Fe/H]}$=0.2 dex (“Model smooth”). The analytical fit to this distribution, indicated as “smooth (fit)” in the likelihood definition of eq. \ref{like}, is also shown. }
\label{fig_MDF}
\end{figure*}

\subsection{Nucleosynthesis prescriptions}
\label{nucleo_sec}
For both the inner Galactic disc and the NSD,  we adopt the same set of stellar yields proposed by \citet[][model 15]{romano2010}. It is important to stress that this set has also been used for the  modelling  of  the Galactic bulge in \citet{matteucci2019} and more recently to explain the inner bulge chemical abundances new data in the paper of \citet{nieuw2023}.
For low- and intermediate-mass stars (0.8–8 M${\odot}$) we adopt the metallicity–dependent yields of \citet{karakas2010}, which include the contribution of thermal pulses.
For massive stars (with masses $\gtrsim 11$–13 M${_\odot}$, depending on the explosion energy), which give rise to either Type II supernovae or hypernovae, we adopt metallicity–dependent He, C, N and O yields computed with the Geneva evolutionary models that include the combined effects of stellar rotation and mass loss \citep{meynet2002, hirschi2005, hirschi2007, ekstrom2008}.
For the elements heavier than oxygen, we use the stellar yields calculated by \citet{koba2006}.

\section{Data sample for the NSD metallicity distribution}
\label{sec_data_MDF}
The models presented in this work will be constrained by the MDF derived from an updated stellar sample of candidate NSD members originally presented by \citet{Schultheis2021}. 
In \citet{Schultheis2021}, they presented the  kinematics and global metallicities for  NSD candidates  based on the observations of K/M giant stars via a dedicated
KMOS (VLT, ESO) spectroscopic survey.

We call Sample A the same subsample of stars used in \citet{sormani2022} to construct self-consistent dynamical models of the NSD. These authors used a preliminary version of the second version of Via Lactea survey Infrared Astrometric Catalogue \citep[VIRAC2,][]{smith2025} to exclude stars that had large error on proper motions. Here, we do not use the proper motions, but we simply select the same sample of stars for consistency.
By simultaneously modelling the contamination from the Galactic bar with an $N$-body simulation, they were able to assign posterior membership probabilities to individual stars, thereby providing a robust probabilistic framework for the identification of genuine NSD members. 
The Sample A comprises 1806 stars. In Fig.~\ref{fig_MDF}, the MDF in terms of [Fe/H] for this sample is shown as filled pink histograms.  
As anticipated in Section~\ref{NSD_sec}, the MDF of the selected NSD stars covers a wide metallicity range, from $-1.44$~dex to $0.95$~dex in [Fe/H], with a median value of [Fe/H]$_{\rm median} = 0.08$~dex.
 In Sample A, we find that $\sim22$\% of the stars have total velocities exceeding $v_{\mathrm{tot}} > 250 \,\mathrm{km \, s^{-1}}$. These high velocities are characteristic of bulge populations, and the derived fraction should be regarded as a conservative lower limit.  

 Hence, we also considered another sample to  minimise the contamination of NSD disc stars by the Galactic bulge. From Figure 13 of \citet{Schultheis2021}, it can be seen that for metallicities below [M/H] $<$ –0.3/–0.5 dex, the correlation between metallicity and velocity dispersion overlaps with that of the Galactic bulge, indicating a high level of bulge contamination. To account for this possibility, we defined a second subsample starting from Sample A by excluding  all stars with [Fe/H] < –0.3 dex (Sample B).  Sample B is composed by 1544 stars.


\section{Fitting procedure with MCMC methods} \label{fitting}
Here,  we present the main characteristics of the Bayesian analysis based on MCMC methods adopted in this work. In Section \ref{sub_like}, we introduce the adopted likelihood, and in Section \ref{sec_priors} we introduce  the parameters priors and the MCMC sampling procedure.
\subsection{Likelihood definition}
\label{sub_like}

 Bayesian analysis based on MCMC methods has transformed scientific research in the past decade. 
Since there already exist several textbooks and reviews on Bayesian statistics 
\citep[see e.g.][]{jaynes2003,gelman2013} and on MCMC methods 
\citep[see e.g.][]{brooks2011,sharma2017,hogg2018,speagle2019}, here we only summarise the general framework 
and highlight the aspects specific to the problem at hand. In the context of Galactic chemical evolution, 
Bayesian inference and MCMC techniques have already proven extremely effective in constraining the 
parameter space of chemical evolution models for the Galactic disc \citep{Spitoni2020,spitoni2021} and dwarf galaxies \citep{cote2017,johnson2023,olcay2025}, motivating their use for the 
present NSD study.

In parameter estimation, Bayes’ theorem provides a way to update the probability of the model parameters 
based on newly available data. It enables the calculation of the posterior probability distribution of the 
parameters given the data,  
\begin{equation}
 P({\bf \Theta}|{\bf x}) = \frac{ P({\bf \Theta}) }{ P({\bf x}) } P({\bf x}|{\bf \Theta}),
 \label{eq:bayes}
\end{equation}
where ${\bf x}$ represents the set of observables, ${\bf \Theta}$ the set of model parameters, 
$P({\bf x}|{\bf \Theta}) \equiv \mathscr{L}$ the likelihood (i.e. the probability of observing the data 
given the model parameters, $P({\bf \Theta})$ the prior (i.e. the probability of the model parameters 
before considering the data), and $P({\bf x})$ the evidence (i.e. the total probability of the data under 
all possible parameter values). The evidence is a normalising constant obtained by integrating the 
likelihood over the full parameter space.

In the present work, the observables are given by the spectroscopic data  already described in Section \ref{sec_data_MDF}  for  the NSD, e.g. 
${\bf x} = \{ [{\rm Fe}/{\rm H}]\}$, 
while the free parameters are those defining the chemical evolution models, 
${\bf \Theta} = \{\tau, \ \nu, \ \omega\}$, namely the infall timescale, star formation efficiency, 
and wind loading factor, respectively. These parameters are explored using MCMC sampling, and their 
posterior distributions are constrained by the observed MDF and abundance-ratio trends of NSD stars.

To identify the best-fitting chemical evolution models  we adopt a Bayesian framework employing MCMC techniques, following the methodology outlined in \citet{Spitoni2020} applied to the MDFs this time \citep{cescutti2020}. Below, we summarise the main elements of the fitting procedure; further methodological details can be found in the aforementioned works.

To evaluate the likelihood $\mathscr{L}$ of a given chemical evolution model with respect to the observed metallicity distribution function (MDF), we proceed as follows. 
We apply a Gaussian smoothing to the normalised predicted  MDF to account for observational uncertainties $(\text{MDF}_{\rm Model \, smooth})$. The width of the smoothing kernel is set to $\sigma_{\rm [Fe/H]}$=0.2 dex, consistent with the sample errors.
The predicted  $\text{MDF}_{\rm Model \, smooth}$ is then interpolated using a one-dimensional linear spline function ($\text{MDF}_{\rm Model \, smooth (fit)}$).  The observational data consist of individual stellar [Fe/H] measurements. The log-likelihood is then calculated as the sum of the logarithms of the interpolated MDF evaluated at each observed metallicity:

\begin{equation}
\label{like}
\log \mathscr{L} = \sum_{i=1}^{N} \log \left[ \text{MDF}_{\rm Model \, smooth(fit)} \left( [\text{Fe/H}]_{\text{obs.}, i} \right) \right],
\end{equation}
where $N$ is the number of stars in the observed sample. This log-likelihood function is used within the MCMC framework to constrain the posterior distribution of the model parameters.

\subsection{Parameter priors \& MCMC Sampling Procedure }
\label{sec_priors}
Uniform priors are adopted for all model parameters. 
The infall timescale, $\tau$, is allowed to vary within $0.1 < \tau < 10$ Gyr. 
The wind loading factor, $\omega$, which regulates the strength of galactic outflows, 
is sampled uniformly in the range $0 < \omega < 10$. 
The star formation efficiency, $\nu$, is explored over the interval $0.1 < \nu < 100$ Gyr$^{-1}$. 
These intervals ensure a comprehensive exploration of parameter space, 
covering the typical values adopted in chemical evolution models of a wide variety of systems 
—including bulges, galactic discs, and dwarf and irregular galaxies— 
as discussed by \citet{matteucci2021}.

The posterior distributions are sampled using  the aﬃne invariant MCMC ensemble sampler, "\texttt{emcee}: the mcmc hammer" code\footnote{\href{https://emcee.readthedocs.io/en/stable/}{https://emcee.readthedocs.io/en/stable/};
\href{https://github.com/dfm/emcee}{https://github.com/dfm/emcee}}, proposed by  \citet{goodman2010,foreman}. This method allows for efficient exploration of high-dimensional parameter spaces and is well-suited for problems involving complex, multi-modal distributions. We initialised the chains with 14 walkers and
ran the sampler for 2800 steps. The computations were performed on the LEONARDO DCGP (Data Centric General Purpose) of  CINECA, using 42 cores and 24 GB of memory.

\begin{figure*}
    \centering
    \includegraphics[scale=0.42]{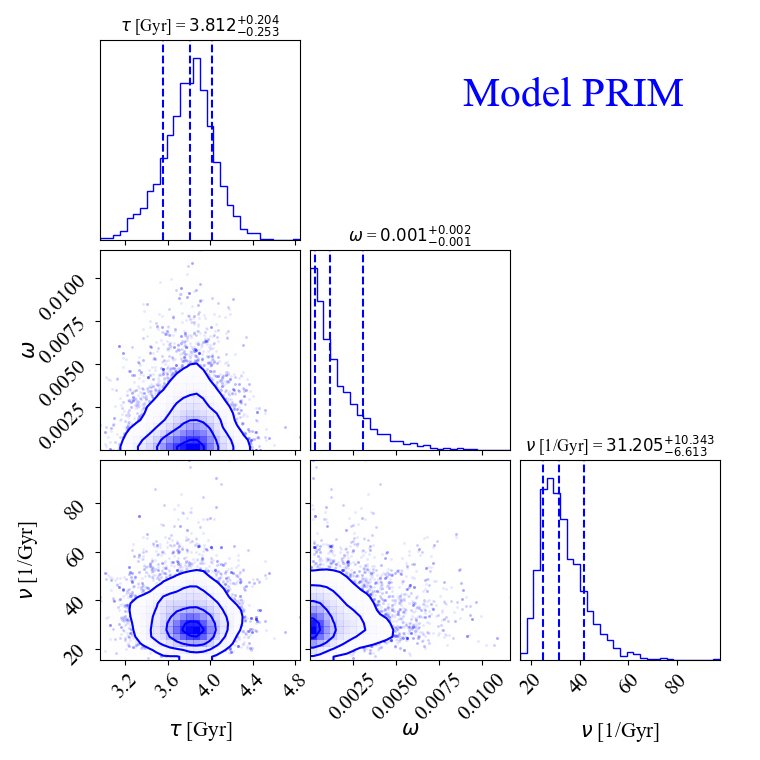}
     \includegraphics[scale=0.42]{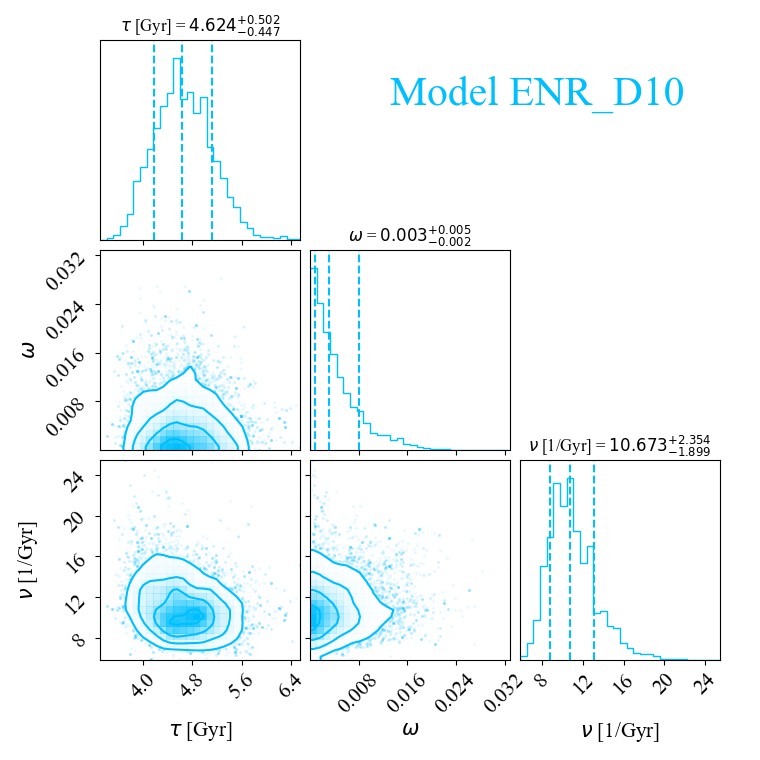}
 \includegraphics[scale=0.42]{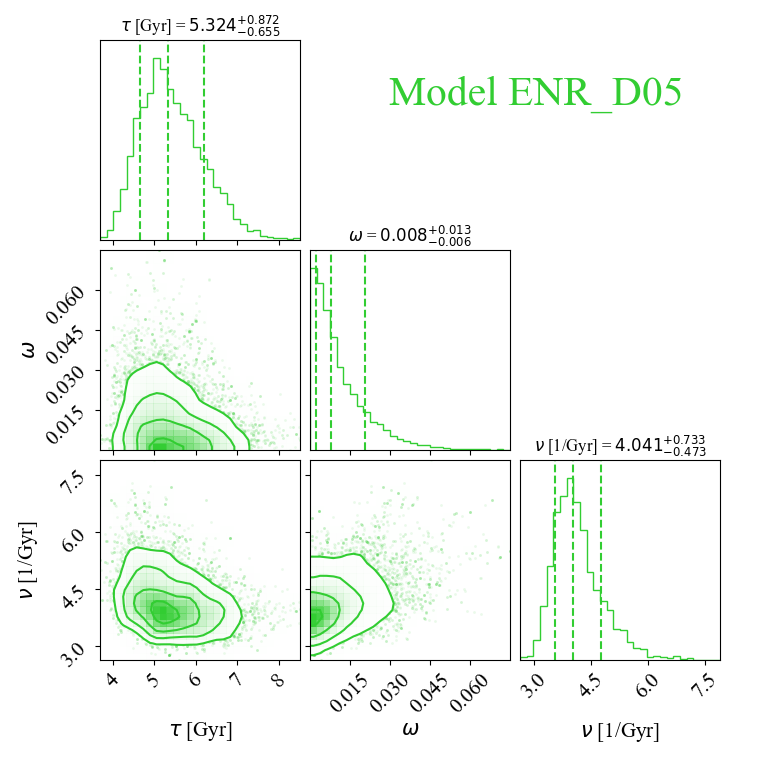}
 \includegraphics[scale=0.42]{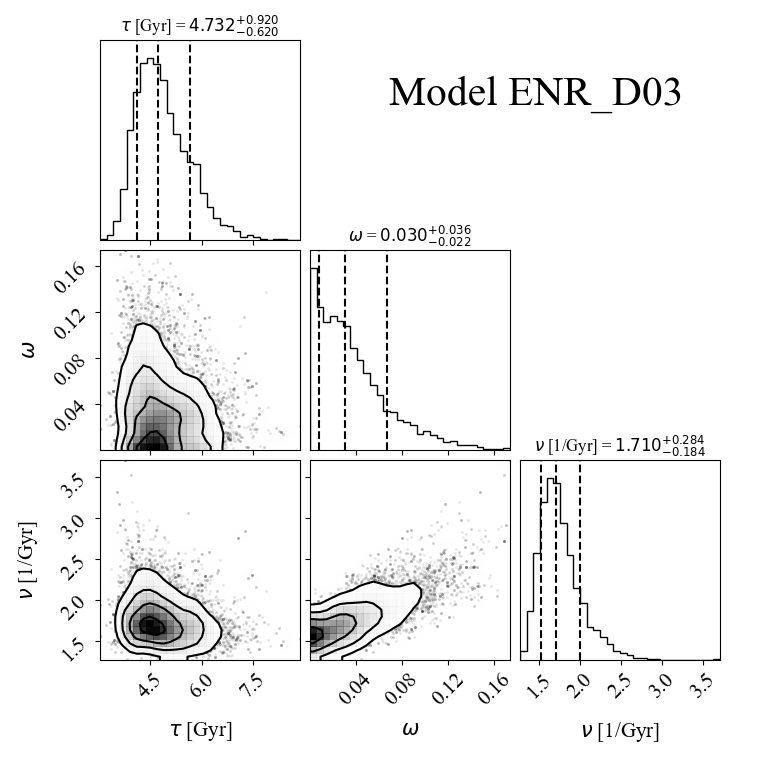}
 \caption{Corner plots showing the PDFs of the chemical–evolution model parameters for the different models reported in Table \ref{tab_mcmc} (one in each panel).
For each parameter,
the median and the 16th and 84th percentiles of the posterior PDF
are shown with dashed lines above the marginalised PDF. All models have adopted the Sample A data.}
\label{corner}
\end{figure*}

\begin{figure}
\centering
 \includegraphics[scale=0.38]{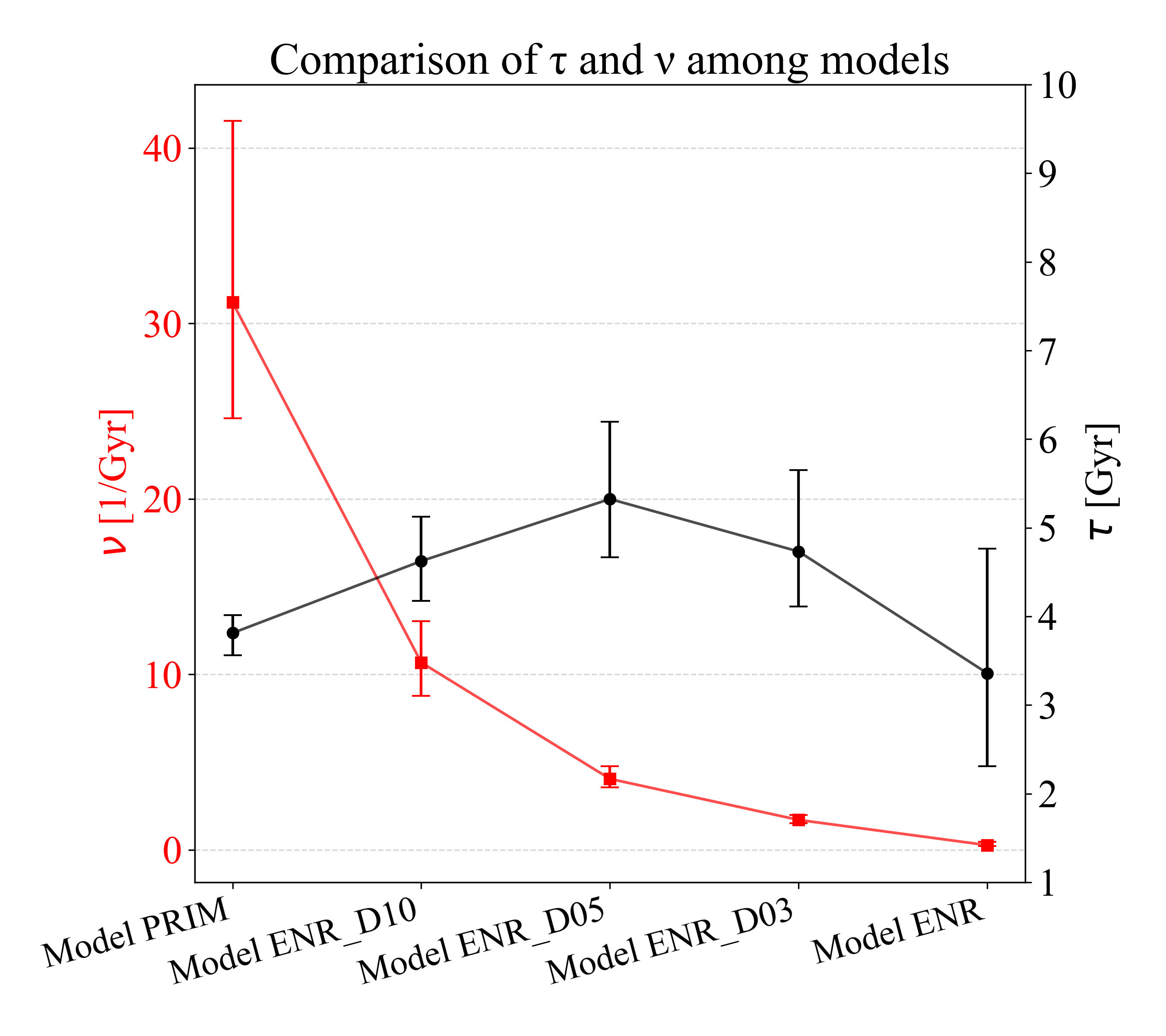}
         \caption{Comparison of the infall time‐scale $\tau$ (red symbols and line, left‐hand y‐axis) and the star‐formation efficiency $\nu$ (black symbols and line, right‐hand y‐axis) for the different models reported in Table \ref{tab_mcmc}. }
\label{tau_nu}
\end{figure}

\begin{figure}
\centering
 \includegraphics[scale=0.38]{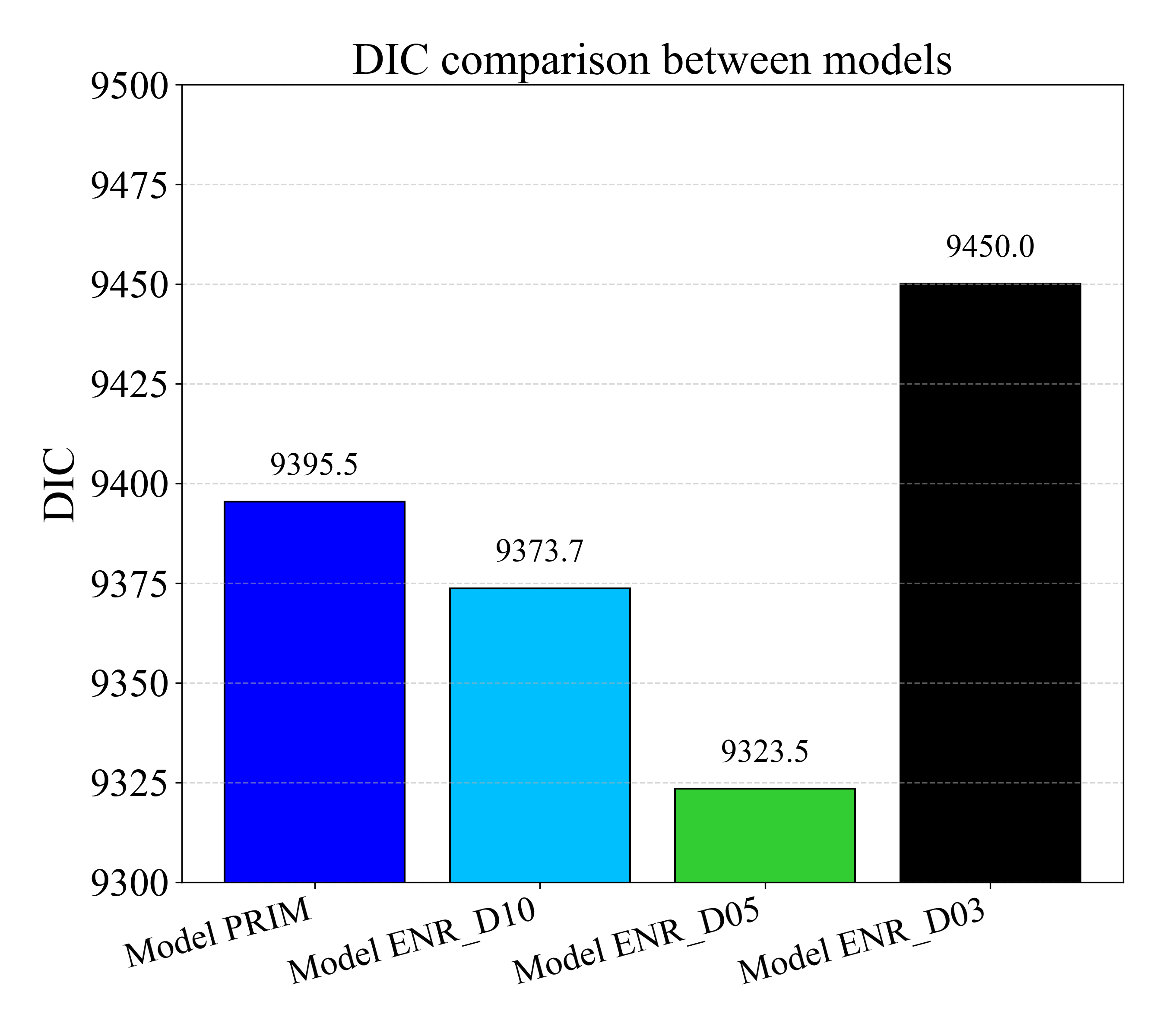}
         \caption{Deviance Information Criterion (DIC) comparison between models in presence of  gas dilution for the NSD. }
\label{DIC}
\end{figure}

\begin{figure}
    \centering
    \includegraphics[scale=0.42]{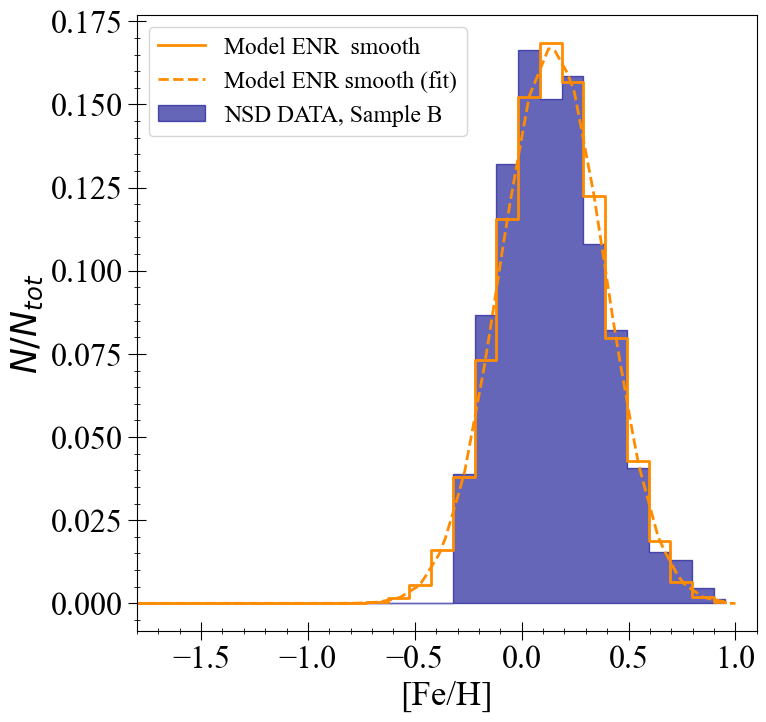}
     \includegraphics[scale=0.44]{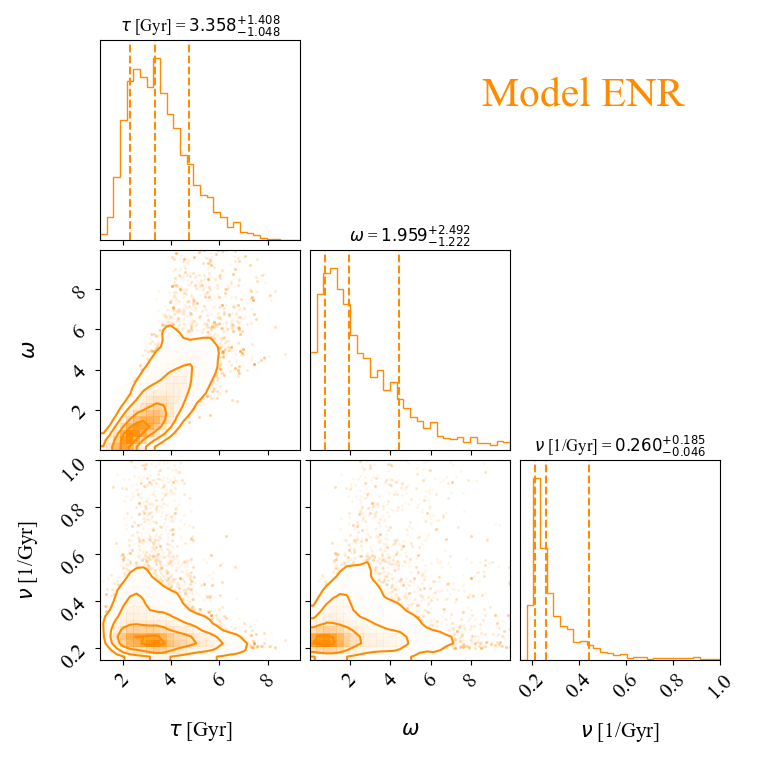}
 \caption{  Results of the Model ENR. {\it Upper panel}: As in  Fig. \ref{fig_MDF} but compared to the MDF obtained by using the Sample B 
(filled purple distribution, see Section \ref{sec_data_MDF} for details) and the predictions of the best-fit Model ENR (with parameters listed in Table \ref{tab_mcmc}). {\it Lower panel}: As in Fig. \ref{corner} but for Model ENR.}
\label{corner_ENR}
\end{figure}

\section{Results for the NSD chemical evolution}
\label{sec_results}
 In this Section, we present the main results of our chemical evolution models for the NSD, obtained within the Bayesian MCMC framework described in Section~\ref{fitting}, and constrained by the MDFs derived from Sample A and Sample B, as introduced in Section~\ref{sec_data_MDF}. In Section~\ref{sec_results_cem_comp}, we first present the different models characterised by different degrees of chemical dilution in the inflowing gas from the inner disc.
In Section~\ref{sec_results_MCMC}, we  discuss the best-fit model parameters as  inferred from the MCMC analysis. In Section~\ref{results_others}, we then compare these best-fit models with additional observational constraints, including the star formation history estimated by \citet{sanders2024}. Finally, in Section~\ref{results_ryde}, we confront the model predictions with abundance ratios as reported by \citet{Ryde2025}.

\subsection{Chemical composition of the inflowing gas into the NSD}
\label{sec_results_cem_comp}
The observed MDF of the Sample A in Fig. \ref{fig_MDF} exhibits a significant sub-solar metallicity tail, extending from [Fe/H] $\sim -1.2$ to $0$~dex. In contrast, the chemical composition of the inner Galactic disc at the time of bar formation ($\sim 8$ Gyr ago), as presented in Section \ref{model_4kpc}, is already close to solar value ([Fe/H]=-0.10 dex). Therefore, in this case, a scenario in which the NSD is built solely from disc gas cannot account for the sub-solar metallicity component of the MDF.

In the Bayesian framework (see Section \ref{fitting}), the MDF will be  used as observational data to be reproduced by the model. This requires considering a certain level of dilution in the inflowing gas. In this study, we investigate different levels of chemical enrichment for ${X}_{ \mathscr{F}, NSD, \, i}(t)$ in order to assess which degree of dilution provides the most consistent reproduction of the observed metallicity distribution of  stars in the NSD.

Using as a constraint the NSD stars of the  Sample A, we explore four different prescriptions for the chemical composition of the gas inflowing into the NSD, ${X}_{ \mathscr{F}, NSD, \, i}(t)$, and summarised as:
\definecolor{deepskyblue}{RGB}{0,191,255}
\definecolor{limegreen}{RGB}{50,205,50}

\begin{equation}
{X}_{ \mathscr{F}, NSD, \, i}(t) =
\begin{cases}
\text{Primordial}, & \text{Model PRIM}  \\
X_{4D,i}(t)/10, & \text{Model ENR\_D10}  \\
X_{4D,i}(t)/5, & \text{Model ENR\_D05}  \\
X_{4D,i}(t)/3, & \text{Model ENR\_D03}. \\
\end{cases}
\end{equation}

In the first case (Model PRIM), the accreting gas is assumed to be primordial, representing an extreme scenario. In the other three models, the inflow is considered to be partially enriched in a self-consistent manner by the inner Galaxy as a function of evolutionary time, but only for epochs successive to the formation of the bar. Specifically, the enrichment level is scaled down by a factor of 10 (ENR\_D10), 5 (ENR\_D05), or 3 (ENR\_D03) with respect to the chemical composition of the disc gas, $X_{4D}(t)$. These prescriptions enable us to explore the effect of different degrees of pre-enrichment on the resulting metallicity distribution of the NSD.

The fact that the gas forming the NSD should be less metal rich than the gas coming from the inner disc, can find an explanation in the addition of more metal poor gas originating from the early formation of the bulge or thick disc \citep[see also][]{Ryde2025}. 
Moreover, in their recent study, \citet{sextl2025} analysed the nuclear star-forming rings of four disc galaxies (NGC 613, NGC 1097, NGC 3351, and NGC 7552). By separating the spectral contributions of young and old stellar populations, they identified a wide range of ages and metallicities among the oldest stars in these nuclear rings. This diversity suggests a continuous inflow of metal-poor gas and recurrent episodes of star formation occurring over several Gyr.

Finally, we performed an additional test using the MDF of Sample B (see Section \ref{sec_data_MDF}), excluding stars with [Fe/H] < –0.3 dex. In this specific case, we adopted a NSD model in which the chemical composition of the accreted  gas is assumed to be identical to that of the inner Galactic disc as a function of  time, i.e. ${X}_{ \mathscr{F}, NSD,i}(t) = X_{4\mathrm{D},i}(t)$. We refer to this model as Model ENR.

\subsection{Results of the MCMC analysis}
\label{sec_results_MCMC}
 Figure~\ref{corner} displays the posterior probability density functions (PDFs) of the three free parameters of our chemical evolution models for the NSD, obtained using Sample A as the observational constraint: the infall timescale ($\tau$), the wind loading factor ($\omega$), and the star formation efficiency ($\nu$). Figure \ref{corner} was generated using the final 800 iterations of the chains, after discarding the burn-in phase. With a mean autocorrelation time  of  $\tau_{emcee} \sim$ 48 for the four models considered, the 2800 iterations performed were sufficient to ensure chain convergence.
The median values and $1\sigma$ confidence intervals are summarised in Table~\ref{tab_mcmc}.

\definecolor{deepskyblue}{RGB}{0,191,255}
\definecolor{limegreen}{RGB}{50,205,50}


 \begin{table*}[]
\begin{center}
\tiny
\caption{
Main properties of our best-fit models for the Nuclear Stellar Disc varying the chemical enrichment of the gas flow. 
}
\label{tab_mcmc}
\begin{tabular}{|c|ccccc|}
\hline
  \hline
 &  \multicolumn{5}{|c|}{\it NSD Models}   \\
 &  &&&&\\
  &  \textcolor{blue}{PRIM}&\textcolor{deepskyblue}{ENR\_D10} &\textcolor{limegreen}{ENR\_D05}& ENR\_D03&\textcolor{orange}{ENR}\\
  &  &&&&\\ 
\hline
 &  &&&&\\
${X}_{ \mathscr{F}, NSD}(t)$ &Primordial&${X_{4D}}(t)/10$ &${X_{4D}}(t)/5$&${X_{4D}}(t)/3$  &${X_{4D}}(t)$    \\
&&&&&\\

\hline
&&&&&\\
 &  \multicolumn{5}{c|}{\it  MCMC Results}   \\
 &&&&&\\
  \hline
 &&&&&\\
 $\tau$ [Gyr]&3.812$^{+0.204}_{-0.253}$&4.624$^{+0.502}_{-0.447}$& 5.324$^{+0.872}_{-0.655}$&4.732$^{+0.920}_{-0.620}$& 3.358$^{+1.408}_{-1.048}$\\

&&&&&\\
$\omega$ &0.001$^{+0.002}_{-0.001}$ &0.003$^{+0.005}_{-0.002}$& 0.008$^{+0.013}_{-0.006}$ & 0.030$^{+0.036}_{-0.022}$&1.959$^{+2.492}_{-1.222}$ \\

&&&&&\\
 $\nu$ [Gyr$^{-1}$] &31.205$^{+10.343}_{-6.613}$&10.673$^{+2.354}_{-1.899}$&  4.041$^{+0.733}_{-0.473}$&1.710$^{+0.284}_{-0.184}$& 0.260$^{+0.185}_{-0.046}$\\
 &&&&&\\
   \hline
&  \multicolumn{4}{c|}{} &   \\
{  NSD  DATA}&  \multicolumn{4}{c|}{Sample A} & Sample B  \\
 \hline
\end{tabular}
\end{center}
\tablefoot{In the upper part of the Table we show the chemical composition
of the gas flow into the NSD.  In the lower one, we indicate 
 the  accretion timescale $\tau$, loading factor  $\omega$ and star formation efficiency $\nu$  from our MCMC  estimates for the four considered models. In the last row, we report the observed samples adopted as constraints in the MCMC analysis.}
\end{table*}


A consistent outcome across all models is that the NSD wind loading factor is very low, with typical values  around $\omega \simeq 0.01$. This indicates that outflows have played a negligible role in shaping the chemical evolution of the NSD, consistent with its deep gravitational potential in which it is embedded.  

The inferred infall timescales lie in the range $\tau \simeq 4$--5 Gyr and show little dependence on the assumed chemical composition of the inflowing gas (Fig. \ref{tau_nu}, black points). 
These values are consistent with predictions of the thin  disc component in the inner Galactic regions  \citep{chiappini2001,palla2020,spitoni2021} 
By contrast, the star formation efficiency shows a strong dependence on the adopted enrichment prescription (Fig. \ref{tau_nu}, red points). Models with primordial or highly diluted inflows (PRIM and ENR\_D10) require very high SFEs ($\nu \gtrsim 10$ Gyr$^{-1}$), comparable to those associated with the secular evolution of the bulge \citep{matteucci2019, molero2024}, in order to reproduce the observed MDF. In cases with moderate dilution (ENR\_D05 and ENR\_D03), the inferred values are instead closer to those typical of disc evolution in the inner Galactic regions ($\nu \lesssim 5$ Gyr$^{-1}$).

To assess the relative quality of the chemical evolution models,  we adopt the Deviance Information Criterion (DIC), which provides a balance between the quality of the fit and the effective model complexity. In practice, the DIC was computed directly from the MCMC chains of the posterior probability. For each sampled parameter vector $\theta$, we evaluated
\begin{equation}
    \chi^{2}({\bf \Theta}) = -2 \log P({\bf \Theta}|{\bf x}).
\end{equation}
From these values, the mean deviance and the minimum deviance were derived across the chain. The DIC was then estimated using the following expression:
\begin{equation}
    {\rm DIC} = 2 \, \langle \chi^{2} \rangle - \min(\chi^{2}),
\end{equation}
where $\langle \chi^{2} \rangle$ denotes the mean over the posterior distribution and $\min(\chi^{2})$ is the best-fit value found in the chain. 
The resulting DIC values allow a direct comparison between the inflow scenarios considered in this work, with lower values indicating the statistically preferred model.

The relative performance of the four models is illustrated in Fig. \ref{DIC}, which shows their predicted DIC values. Model ENR\_D03  provides the poorest fit, as indicated by its highest DIC value. This result is consistent with Fig.~\ref{fig_MDF}, where Model ENR\_D03  clearly under-predicts stars in the low-metallicity tail of the MDF.
\begin{figure}
\centering
 \includegraphics[scale=0.38]{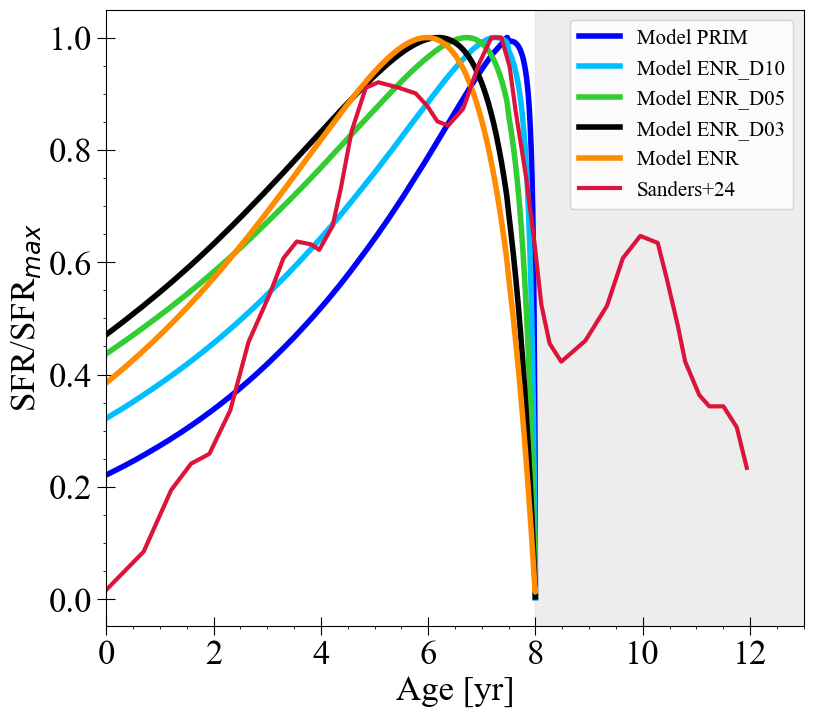}
           \includegraphics[scale=0.38]{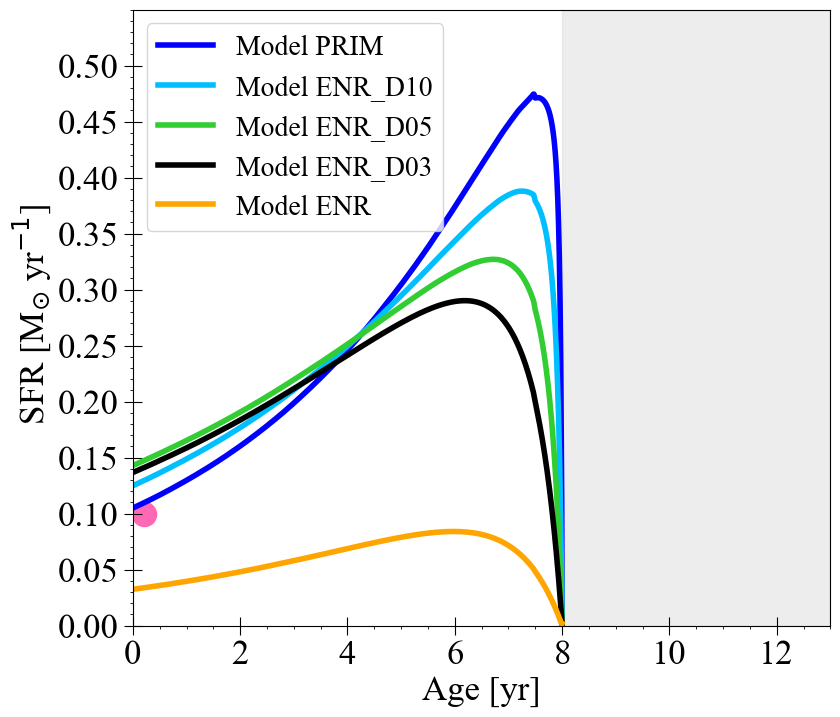}
 \caption{Temporal evolution of the star–formation rate (SFR) predicted by the NSD models listed in Table \ref{tab_mcmc}.
{\it Upper panel}: SFR normalised to its maximum value and compared with the recent results of \citet{sanders2024}.
{\it  Lower panel}: SFR in physical units of M$_{\odot}$ yr$^{-1}$. With the magenta point is indicated  the  \citet{henshaw2023} estimate of 0.1 M$_{\odot}$ yr$^{-1}$.
In both panels, the grey shaded region marks the epoch preceding the formation of the Galactic bar. }
\label{SFR_fig}
\end{figure}
\begin{figure}
\centering
 \includegraphics[scale=0.38]{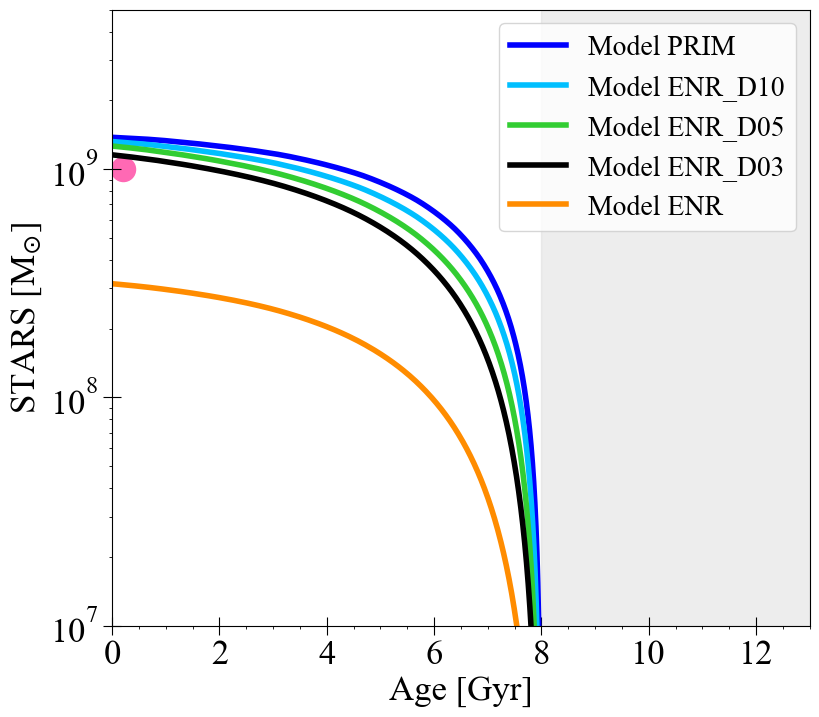}
         \caption{As in Fig. \ref{SFR_fig} but for the evolution of the total stellar mass predicted by the models. The magenta point indicates the estimated present-day the total stellar mass  for NSD  of $10^9$ M$_{\odot}$ \citep{Launhardt2002, NoguerasLara2020,schultheis2025}. }
\label{fig_stars}
\end{figure}

\begin{figure}
\centering
 \includegraphics[scale=0.38]{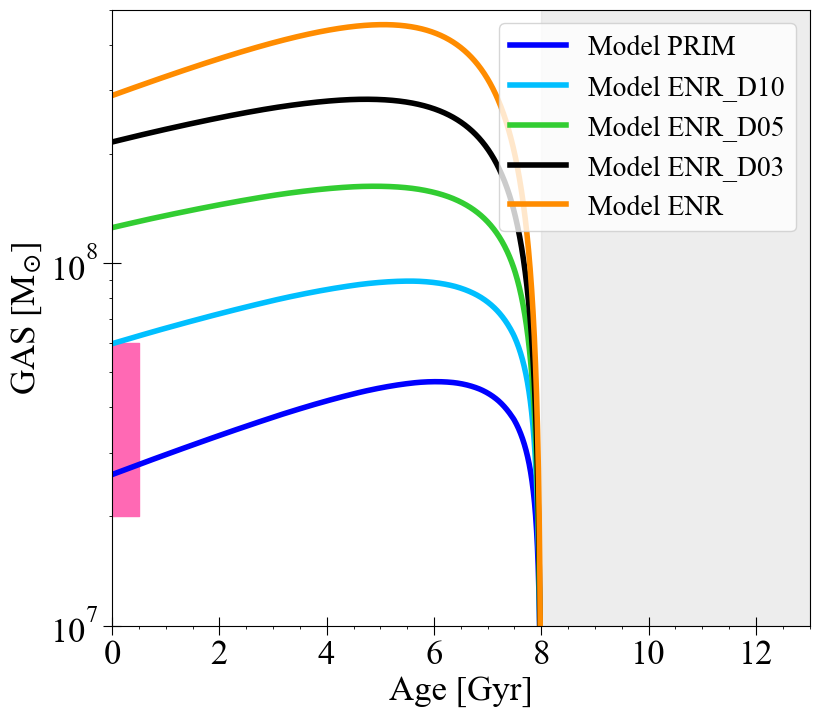}
         \caption{Same as Fig.~\ref{SFR_fig}, but showing the evolution of the gas mass predicted by the models. With the pink area we indicate the range of the estimated mass of the molecular gas of the CMZ  \citep{dahmen1998,ferriere2007}. }
\label{fig_mass_gas}
\end{figure}

The next least favoured model is PRIM, which significantly overproduces stars in the metallicity range $-0.5 \lesssim \mathrm{[Fe/H]} \lesssim 0$, in clear disagreement with the observed MDF. By contrast, the best-fit model is ENR\_D05, which achieves the lowest DIC value while accurately reproducing both the low-metallicity tail and the peak of the MDF. ENR\_D10 offers  an improvement over the primordial case.
Taken together, these results indicate that the flowing gas that built the NSD must have been diluted relative to the inner disc composition by a factor of about larger than 3 and smaller than 10. Pure inner-disc inflow is firmly ruled out.  
 This suggests that, in addition to bar-driven inflows from the inner disc, lower-metallicity gas—possibly originating from the thick disc or from recent infall of pristine gas \citep{sextl2025} — also contributed to the NSD’s build-up.

As discussed in Section \ref{sec_results_cem_comp}, in Model ENR we also explore the case where the chemical composition of the inflowing gas is identical to that of the inner disc. This configuration is compatible only to Sample B, where stars with [Fe/H] < –0.3 dex were excluded.
For this analysis, we narrowed the prior parameter space for the galactic outflow strength and star formation efficiency, guided by  the best-fit results obtained for Sample A, restricting the ranges to $0 < \omega < 10$ and $0.1 < \nu < 5$ Gyr$^{-1}$.
Figure \ref{corner_ENR} shows that, in this case, the wind loading factor of the best-fit model is no longer negligible, with $\omega = 1.959^{+2.492}_{-1.222}$. Conversely, the best-fit star formation efficiency is quite low, $\nu = 0.260^{+0.185}_{-0.046}$ Gyr$^{-1}$. This outcome is consistent with the results obtained for Sample A, enforcing a chemically enriched inflow that drives the system toward reduced star formation, as a consequence of both stronger galactic winds and lower star formation efficiency.
 We emphasise that, in this study, Sample B should be regarded as an extreme case, presented primarily to illustrate the only scenario in which an NSD model can be formed from gas inflow with inner-disc chemical composition, without any dilution.

 We already discussed that the low-metallicity gas required to dilute the inflow in the inner disc could originate either from the thick disc or from recent external accretion events. These two scenarios may be distinguished observationally through detailed chemical abundance patterns.
The [Mg/Mn] versus [Al/Fe] diagram \citep{hawkins2015,fernandes2023,vasini2024}  offers, in principle, a useful tool for distinguishing between these two scenarios  because sensitive to  different star formation histories. This sensitivity arises from the distinct nucleosynthetic origins and enrichment timescales of the elements involved. Mn is mainly released on long timescales by Type Ia SNe, which are also responsible for most of the iron production. In contrast, Mg and Al are primarily synthesised in massive stars and expelled into the interstellar medium by core-collapse SNe, which contribute only a minor fraction of the total iron budget.
Thick-disc stars are known to exhibit enhanced [Al/Fe] and [Mg/Mn] ratios relative reflecting their dominant enrichment by core-collapse supernovae. On the other hand, gas accreted from the circumgalactic medium or from gas-rich satellites is expected to show lower [Al/Fe] and distinct [Mg/Mn] signatures due to slower chemical evolution and a larger contribution from Type Ia SNe. Future observations with MOONS will enable precise measurements of these  abundance ratios and thus offer a valuable opportunity to distinguish between Galactic (thick-disc) and external (accretion-driven) origins of the metal-poor inflow. However, as emphasised by \citet{vasini2024}, the [Al/Fe]–[Mg/Mn] diagram remains theoretically uncertain. Its effectiveness as a diagnostic of star formation histories will critically depend on advances in stellar nucleosynthesis modelling for these elements, whose production mechanisms are still subject to substantial uncertainties.

\subsection{Comparison with other observational data}
\label{results_others}

\begin{figure}
\centering
 \includegraphics[scale=0.38]{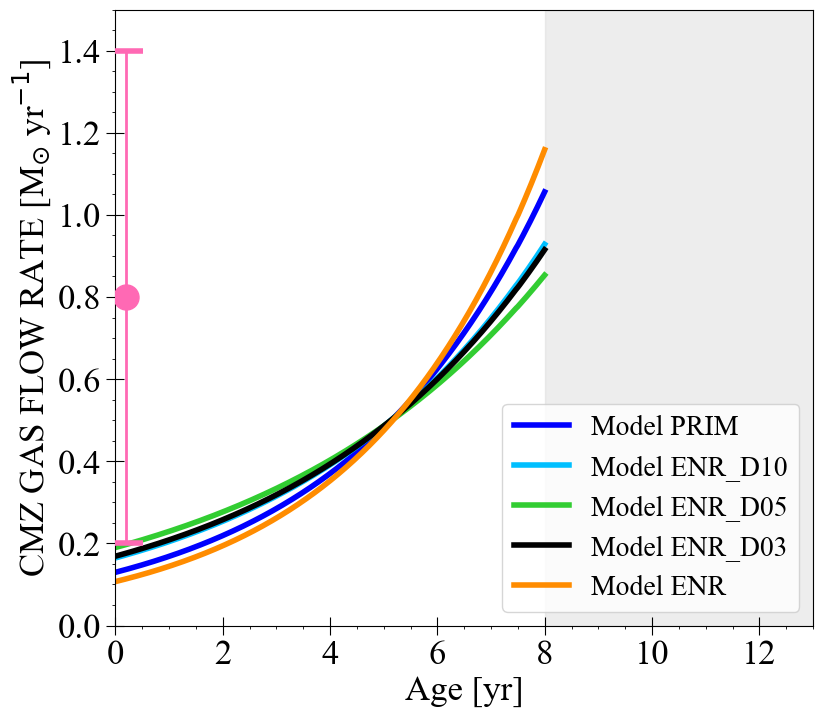}
         \caption{Temporal evolution of the inflow gas rate   predicted by the models extended for all the region of the CMZ, assuming the same total surface mass density of eq. (\ref{sigma}) and multiplied by an area of $\pi R_{CMZ}^2$ where $R_{CMZ}$=150 pc. The pink point and associated error bars stands for the estimated value by \citet{hatchfield2021} using hydrodynamic simulations with AREPO. }
\label{fig_flow}
\end{figure}

In addition to the MDF of NSD stars presented in Section~\ref{sec_data_MDF}, our models can also be compared with further observational constraints that are not directly included in the MCMC analysis. 
Figures~\ref{SFR_fig}--\ref{fig_flow} illustrate the predicted temporal evolution of the star formation rate, stellar mass, gas mass, and gas inflow rate for the four enrichment scenarios.

 \begin{figure*}
\centering
 \includegraphics[angle=0, scale=0.41]{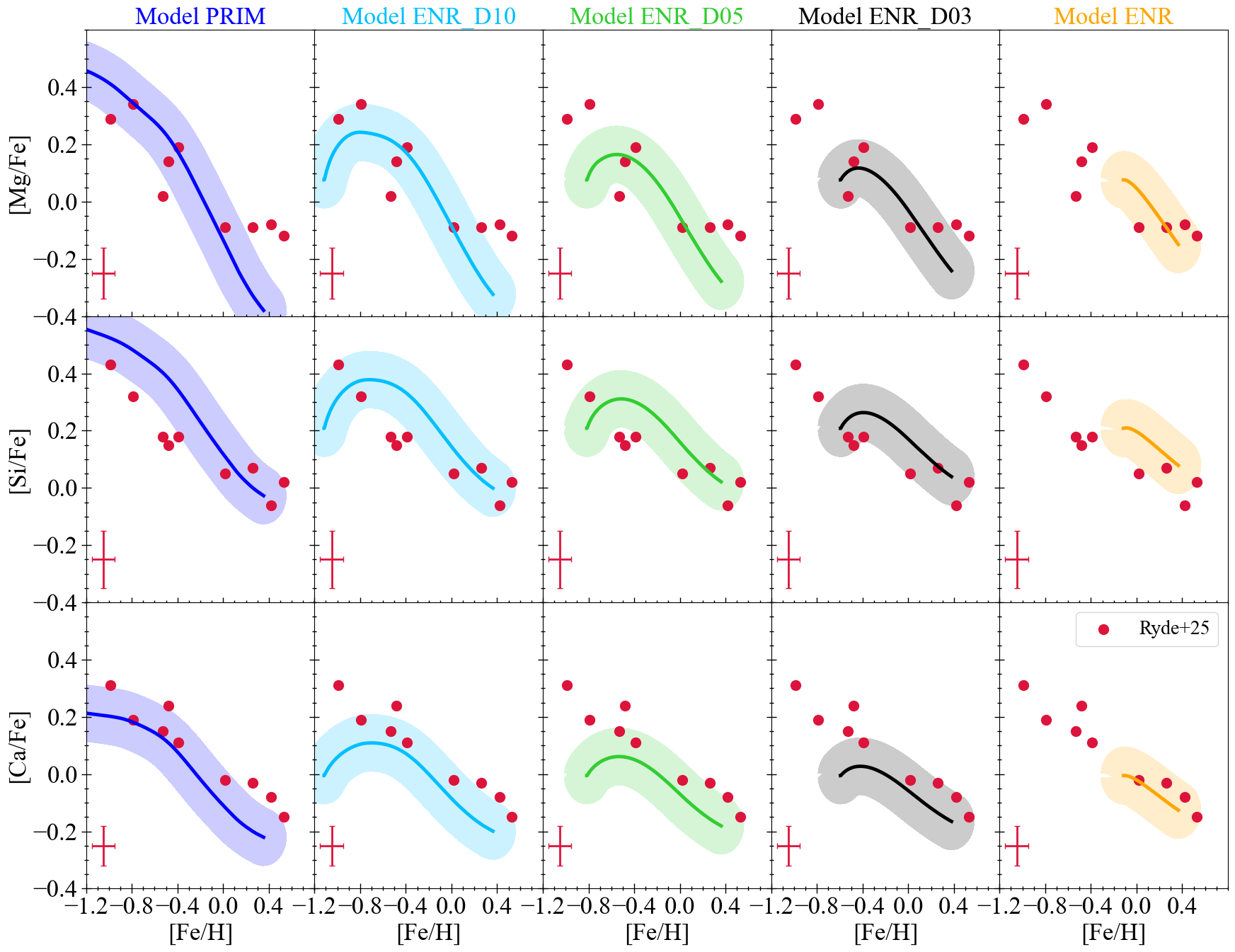}
 \caption{Comparison between model predictions and NSD data from \citet[][red points]{Ryde2025}. Each panel shows the evolution of [Mg/Fe], [Si/Fe], and [Ca/Fe] as a function of [Fe/H], as predicted by the four chemical evolution models considered in this work (PRIM, ENR\_D10, ENR\_D05, and ENR\_D03; see Table~\ref{tab_mcmc} for the adopted parameters). Shaded regions around the model curves indicate the uncertainties, obtained by Gaussian smoothing with $\sigma_{\rm[Fe/H]}=0.2$~dex - consistent with the MDF fitting procedure - and $\sigma_{\rm[X/Fe]}=0.1$~dex, where X denotes the considered element. Typical uncertainties of the \citet{Ryde2025} data are indicated in the lower-left corner of each panel by a red cross.
 }
\label{fig_ryde}
\end{figure*}

The predicted star formation histories in Fig.~\ref{SFR_fig} show a pronounced enhancement of star formation immediately after the formation of the Galactic bar (about 8 Gyr ago), marking the onset of NSD formation. 
Stronger dilution scenarios lead to sharper peaks closer to this epoch, since a larger number of stars must form rapidly in order to reach the chemical enrichment level observed in present-day NSD stars.

The SFR estimated by \cite{sanders2024} and drawn in the upper panel of Fig.~\ref{SFR_fig} shows a rapid decline in recent times, approaching a near-zero value, while all other proposed models exhibit a less pronounced decrease, retaining 20-45\% of their peak levels at present.
It is important to stress that \citet{sanders2024} derived the NSD star formation history from MIRA variables used as age tracers. These evolved stars provide strong constraints on ancient star formation episodes (around $8 \pm 1$ Gyr ago), but are less sensitive to very recent activity. This limitation likely explains the almost negligible star formation at the most recent epochs in their reconstruction, in contrast with our models, which predict a more extended star formation history.

The star formation history inferred by \citet{sanders2024} also shows a modest component prior to the epoch of bar formation. As already highlighted in previous Sections, in our modelling we assume that the NSD formed only after the establishment of the Galactic bar \citep{baba2020, cole2014}. We interpret the small pre-bar component in \citet{sanders2024} reconstruction as likely spurious, or the result of contamination from the surrounding bulge– Galactic bar system.

In the lower panel of Fig. \ref{SFR_fig}, we show the temporal evolution of the SFR, expressed in units of M$_{\odot}$ yr$^{-1}$, as predicted by our models with different dilution levels.  Our results for models including gas dilution (i.e. those constrained by Sample A) are in good agreement with the estimated present-time star formation rate of 0.1 M${\odot}$ yr$^{-1}$ reported by \citet{henshaw2023}. Among these, the PRIM model provides the best agreement, whereas the other cases predict values in the range 0.12–0.14 M${\odot}$ yr$^{-1}$.
Conversely, the model without dilution (Model ENR) constrained by Sample B substantially underestimates the \citet{henshaw2023} value, predicting a rate of only 0.032 M$_{\odot}$ yr$^{-1}$.

The predicted build-up of stellar mass (see Fig.~\ref{fig_stars}) is broadly consistent with estimates of the present-day NSD mass of $\sim 10^{9}\,\mathrm{M}_{\odot}$. Our models  based on the Sample A slightly overestimate this value, with predictions ranging from $1.15 \times 10^{9}\, \mathrm{M}_{\odot}$ (Model ENR\_D03) to $1.35 \times 10^{9}\,\mathrm{M}_{\odot}$ (Model PRIM).  On the other hand, Model ENR, based on the Sample B,  underestimates the present-day stellar mass, predicting a value of  $3.14 \times 10^{8}\,\mathrm{M}_{\odot}$.   

In Fig. \ref{fig_mass_gas}, we also present  the predicted evolution in time of the gas mass by different models. As reference we use the estimated value observed for the CMZ of  $2-6 \times 10^{7}\,M_{\odot}$ \citep{dahmen1998,ferriere2007}. We can notice that only models PRIM and ENR\_D10  fall in this observed range.  Models with lower dilution tend to overestimate the present-day gas mass. In fact, since fewer stars are needed to reproduce the observed MDF as indicated in Fig. \ref{fig_stars} (due to the higher chemical enrichment of the infalling gas), the system is left with more gas.  On the same line, Model ENR, where  no chemical dilution of the inflowing gas is considered,  gives  a value of $2.90 \times 10^{8}\,M_{\odot}$.

Finally, Fig. \ref{fig_flow} shows the temporal evolution of the gas inflow rate predicted by the models, extended to the entire region occupied by the CMZ. This is computed by adopting the same total surface mass density as in Eq. (\ref{sigma}), multiplied by the area $\pi R_{\rm CMZ}^2$ with $R_{\rm CMZ}=150$ pc. The pink point with error bars corresponds to the estimate of \citet{hatchfield2021}, obtained from hydrodynamic simulations with AREPO.  All model predictions lie below the 1$\sigma$ \citet{hatchfield2021} estimate. The best agreement with the data is obtained for Model ENR\_D05.

\subsection{Comparison with \citet{Ryde2025} abundance ratios}
\label{results_ryde}
In Fig. \ref{fig_ryde}, we  compare the predicted [Mg/Fe], [Si/Fe], and [Ca/Fe] versus [Fe/H] abundance ratios with  the recent high-resolution  measurements of NSD giants by \citet{Ryde2025}.
 Concerning all models based on the Sample A (see Section \ref{sec_data_MDF}), it is important to note that while all of them reproduce the overall behaviour of the data, the case with purely primordial infall (PRIM) is physically unrealistic, since some chemical enrichment from the inner disc is expected in the NSD. Consistently, the MCMC analysis already indicated that Model PRIM provides the statistically second poorest fit to the observations. By contrast, the ENR\_D10 and ENR\_D05 models, which also yield  a good agreement with the observed   MDF (see Fig. \ref{DIC}), successfully reproduce the observed decline of [$\alpha$/Fe] with increasing metallicity, thereby supporting the requirement of a significant dilution level in the accreted gas.
Taken together, the MDF of the full data set (Sample A) and abundance-ratio comparisons demonstrate that the NSD cannot be explained by the accretion of chemically evolved inner-disc gas alone. Its formation must have involved a substantial contribution of lower-metallicity material, plausibly originating in the thick disc or the Galactic halo.
 Model ENR, calibrated on the MDF of Sample B  stars, exhibits the smallest  evolution in the [Fe/H]–[$\alpha$/Fe] plane and reproduces only the four super-solar metallicity stars reported by \citet{Ryde2025}. Since \citet{Ryde2025} may be contaminated by bulge stars, this scenario cannot be ruled out, but further analyses with larger datasets will be needed to confirm these results. 

An additional comparison can be drawn with the chemo-dynamical models of \citet{friske2025}. 
In their Figures 6 and 7, the predicted gas-phase abundance-ratio trends exhibit a markedly irregular behaviour, in contrast to the smooth [$\alpha$/Fe]–[Fe/H] sequences observed by \citet{Ryde2025}, which are well reproduced by our models. 
Furthermore, the models of \citet{friske2025} do not generate NSD stars with metallicities below [Fe/H] $\simeq -0.1$ dex, whereas such  metal-poor stars are clearly present in the \citet{Ryde2025} sample and are naturally accounted for by our diluted inflow scenarios.

\section{Conclusions}
\label{conclu_sec}
  In this work, we have developed the first dedicated chemical evolution models of  the Milky Way’s Nuclear Stellar Disc, employing a Bayesian framework to fit the observed MDF derived from various subsamples of the \citet{Schultheis2021} dataset, as described in Section \ref{sec_data_MDF}. 
  
  Our main results can be summarised as follows: 

\begin{itemize}

  \item   For models compared to Sample A (full MDF with the low-metallicity tail), we find that a scenario in which the NSD forms exclusively from gas originating in the inner Galactic disc is inconsistent with the observed MDF.  By the time the Galactic bar formed (∼8 Gyr ago), the metallicity of the inner disc had already reached nearly solar values. As a result, such models cannot account for the sub-solar metallicity wing of the MDF observed in this NSD stellar subsample of \citet{Schultheis2021}.

    \item 
    Hence, a certain level of dilution of the inflowing gas is required. The best-fit models correspond to inflows enriched to roughly one-fifth of the inner disc metallicity (ENR\_D05), with an accretion timescale of $\tau= 5.324^{+0.872}_{-0.655}$ Gyr, a relatively  high star-formation efficiency $\nu= 4.041^{+0.733}_{-0.473}$ Gyr$^{-1}$, and a negligible wind loading factor ($\omega= 0.008^{+0.013}_{-0.006}$).  

    \item The Deviance Information Criterion  comparison clearly favours the ENR\_D10  and ENR\_D05 models (inflows enriched to  one-tenth  and one-fifth of the inner disc metallicity, respectively), while the primordial (PRIM) and weakly diluted (ENR\_D03) models are disfavoured. 

    \item  For Model ENR applied to Sample B (stars with [Fe/H] < –0.3 dex were excluded), assuming that the inflowing gas has the same composition as the inner disc, the best-fit solution implies a significant wind ($\omega = 1.959^{+2.492}_{-1.222}$) and a low star formation efficiency ($\nu= 0.260^{+0.185}_{-0.046}$ Gyr$^{-1}$). This is consistent with Sample A results, showing that chemically enriched inflows tend to suppress star formation through stronger winds and reduced efficiency.

    \item The predicted star formation histories of the best-fit models show a prolonged period of activity after bar formation, in agreement with independent constraints from stellar population studies \citep{sanders2024}.

    \item The models tailored to the  Sample A also successfully reproduce the observed [$\alpha$/Fe]--[Fe/H] trends of \cite{Ryde2025},  whereas the Model ENR (based on Sample B) is able to trace only the four super-solar metallicity stars reported by \citet{Ryde2025}.

\end{itemize}
These results suggest that, in addition to bar-driven inflows enriched by the inner disc, lower-metallicity gas - possibly originating from the  Galactic thick disc or from recent infall of gas  \citep{sextl2025}  - also could have played a significant role in the build-up of the NSD.

Future spectroscopic surveys with large multiplex instruments such as MOONS will vastly increase the available sample of NSD stars with high-quality abundances. Further studies, focusing on stellar kinematics and chemical abundances, are needed to assess bulge contamination in NSD sample stars and hence  to constrain the maximum dilution required by the model to reproduce the observations.
Combined with the Bayesian framework developed here, these data  provide stringent tests of the NSD formation scenario and offer new insights into the assembly of the Milky Way’s central regions.

\begin{acknowledgements}
 The authors are grateful to the referee for the helpful suggestions that contributed to improving the clarity and quality of the manuscript. 
E.S.  thanks R. Ingrao  and A. Vasini
for the very useful discussions.
E.S., F.M. and G.C.  thank I.N.A.F. for the  
1.05.24.07.02 Mini Grant - LEGARE "Linking the chemical Evolution of Galactic discs AcRoss diversE scales: from the thin disc to the nuclear stellar disc" (PI E. Spitoni).  
We thank Leonardo-DCGP supercomputer
as part of the INAF Pleadi 
Call  6.
E.S and G.C. thank I.N.A.F. for the  
1.05.23.01.09 Large Grant - Beyond metallicity: Exploiting the full POtential of CHemical elements (EPOCH) (ref. Laura Magrini). 
F.M. thanks I.N.A.F. for the 1.05.12.06.05 Theory Grant - Galactic archaeology with radioactive and stable nuclei. 
F.M. thanks also support from Project PRIN MIUR 2022 (code 2022ARWP9C) "Early Formation and Evolution of Bulge and HalO (EFEBHO)" (PI: M. Marconi).
M.C.S. acknowledges financial support from the European Research Council under the ERC Starting Grant ``GalFlow'' (grant 101116226) and from Fondazione Cariplo under the grant ERC attrattivit\`{a} n. 2023-3014.
This work was also partially supported by the European Union (ChETEC-INFRA, project number 101008324).  Supported by Italian Research Center on High Performance Computing Big Data and Quantum Computing
(ICSC), project funded by European Union - NextGenerationEU - and National Recovery and Resilience Plan
(NRRP) - Mission 4 Component 2 within the activities of Spoke 3 (Astrophysics and Cosmos Observations).
B.T. acknowledges the financial support from the Wenner-Gren Foundation (WGF2022-0041).
\end{acknowledgements}
\bibliographystyle{aa} 
\bibliography{disk}




\end{document}